# Hybrid Intrusion Detection and Prediction multiAgent System, HIDPAS

Farah Jemili

RIADI Laboratory
Manouba University
Manouba 2010, Tunisia

Montaceur Zaghdoud

RIADI Laboratory
Manouba University
Manouba 2010, Tunisia

Mohamed Ben Ahmed

RIADI Laboratory
Manouba University
Manouba 2010, Tunisia

*Abstract*— **This paper proposes an intrusion detection and prediction system based on uncertain and imprecise inference networks and its implementation. Giving a historic of sessions, it is about proposing a method of supervised learning doubled of a classifier permitting to extract the necessary knowledge in order to identify the presence or not of an intrusion in a session and in the positive case to recognize its type and to predict the possible intrusions that will follow it. The proposed system takes into account the uncertainty and imprecision that can affect the statistical data of the historic. The systematic utilization of an unique probability distribution to represent this type of knowledge supposes a too rich subjective information and risk to be in part arbitrary. One of the first objectives of this work was therefore to permit the consistency between the manner of which we represent information and information which we really dispose. Besides, our system integrates host intrusion detection and network intrusion prediction in the setting of a global anti-intrusions system capable to function like a HIDS (Host based Intrusion Detection System) before functioning like NIPS (Network based Intrusion Prediction System). The so proposed anti-intrusions system permits to combine two powerful tools together to permit a reliable host intrusion detection leading to an as reliable network intrusion prediction. In our contribution, we chose to do a supervised learning based on Bayesian networks. The choice of modeling the historic of data with Bayesian networks is dictated by the nature of learning data (statistical data) and the modeling power of Bayesian networks. However, taking into account the incompleteness that can affect the knowledge of parameters characterizing the statistical data and the set of relations between phenomena, the proposed system in the present work uses for the inference process a propagation method based on a bayesian possibilistic hybridization. The so proposed system is adapted to the modeling of reliability with taking into account imprecision.**

## I. INTRODUCTION

The proliferation of Internet access to every network device, the use of distributed rather than centralized computing resources, and the introduction of network-enabled applications has rendered traditional network-based security infrastructures vulnerable to serious attacks.

*Intrusion detection* can be defined as the process of identifying malicious behavior that targets a network and its resources [1]. *Malicious behavior* is defined as a system or individual action which tries to use or access to computer system without authorization and the privilege excess of those who have legitimate access to the system. The term *attack* can be defined as a combination of actions performed by a malicious adversary to violate the security policy of a target computer system or a network domain [2]. Each attack type is characterized by the use of system vulnerabilities based on some feature values. Usually, there are relationships between attack types and computer system characteristics used by the intruder. If we are able to reveal those hidden relationships, we will be able to predict the attack type.

From another side, an attack generally starts with an intrusion to some corporate network through a vulnerable resource and then launching further actions on the network itself. Therefore, we can define the *attack prediction process* as the sequence of elementary actions that should be performed in order to recognize the attack strategy. The use of distributed and coordinated techniques in attacks makes their detection more difficult. Different events and specific information must be gathered from all sources and combined in order to identify the attack plan. Therefore, it is necessary to develop an advanced attack strategies prediction system that can detect attack strategies so that appropriate responses and actions can be taken in advance to minimize the damages and avoid potential attacks.

Besides, the proposed anti-intrusions system should take into account the uncertainty that can affect the data. The uncertainty on parameters can have two origins [3]. The first source of uncertainty comes from the uncertain character of information that is due to a natural variability resulting from stochastic phenomena. This uncertainty is called variability or stochastic uncertainty. The second source of uncertainty is related to the imprecise and incomplete character of information due to a lack of knowledge. This uncertainty is called epistemic uncertainty. The systematic utilization of an unique probability distribution to represent this type of knowledge supposes a too rich subjective information and risk to be in part arbitrary. The system proposed here offers a formal setting adapted to treat the uncertainty and imprecision, while combining probabilities and possibilities.

In this paper we propose an intrusion detection and prediction system which recognizes an upcoming intrusion and predicts the attacker's attack plan and intentions. In our approach, we apply graph techniques based on bayesian reasoning for learning. We further apply inference to recognize the attack type and predict upcoming attacks. The inference process is based on hybrid propagation that takes into account both the uncertain and imprecise character of information.

## II. RELATED WORK

Several researchers have been interested in using Bayesian network to develop intrusion detection and prediction systems. Axelsson in [5] wrote a well-known paper that uses the Bayesian rule of conditional probability to point out the implications of the *base-rate fallacy* for intrusion detection. It clearly demonstrates the difficulty and necessity of dealing with false alarms [6]. In [7], a model is presented that simulates an intelligent attacker using Bayesian techniques to create a plan of goal-directed actions.

In [8], a naïve Bayesian network is employed to perform intrusion detection on network events. A naïve Bayesian network is a restricted network that has only two layers and assumes complete independence between the information nodes (i.e., the random variables that can be observed and measured). Kruegel in [6] proposed an event classification scheme that is based on Bayesian networks. Bayesian networks improve the aggregation of different model outputs and allow one to seamlessly incorporate additional information.

Johansen in [9] suggested a Bayesian system which would provide a solid mathematical foundation for simplifying a seemingly difficult and monstrous problem that today's Network IDS (NIDS) fail to solve. The Bayesian Network IDS (BNIDS) should have the capability to differentiate between attacks and the normal network activity by comparing metrics of each network traffic sample. Govindu in [10] wrote a paper which discusses the present state of Intrusion Detection Systems and their drawbacks. It highlights the need of developing an Intrusion Prediction System, which is the future of intrusion detection systems. It also explores the possibility of bringing intelligence to the Intrusion Prediction System by using mobile agents that move across the network and use prediction techniques to predict the behavior of user.

## III. INTRUSION DETECTION AND PREDICTION SYSTEM

The detection of certain attacks against a networked system of computers requires information from multiple sources. A simple example of such an attack is the so-called *doorknob* attack. In a doorknob attack the intruder's goal is to discover, and gain access to, insufficiently-protected hosts on a system. The intruder generally tries a few common account and password combinations on each of a number of computers. These simple attacks can be remarkably successful [12]. An Intrusion Detection system, as the name suggests, detect possible intrusions [13]. An IDS installed on a network is like a burglar alarm system installed in a house. Through various methods, both detect when an intruder/attacker/burglar is present. Both systems issue some type of warning in case of detection of presence of burglar/intrusion [10].

There are two general methods of detecting intrusions into computer and network systems: anomaly detection and signature recognition [13,14]. Anomaly detection techniques establish a profile of the subject's normal behavior (norm profile), compare the observed behavior of the subject with its norm profile, and signal intrusions when the subject's observed behavior differs significantly from its norm profile. Signature recognition techniques recognize signatures of known attacks, match the observed behavior with those known signatures, and signal intrusions when there is a match. Systems that use misuse-based techniques contain a number of attack descriptions, or 'signatures', that are matched against a stream of audit data looking for evidence of the modeled attacks. The audit data can be gathered from the network [15], from the operating system [16], or from application log files [6].

IDSs are usually classified as host-based or network-based. Host-based systems use information obtained from a single host (usually audit trails), while network based systems obtain data by monitoring the trace of information in the network to which the hosts are connected [17].

A simple question that will arise is how can an Intrusion Detection System possibly detect every single unknown attack? Hence the future of intrusion detection lies in developing an Intrusion Prediction System [10]. Intrusion Prediction Systems must be able to predict the probability of intrusions on each host of a distributed computer system. Prediction techniques can protect the systems from new security breaches that can result from unknown methods of attacks. In an attempt to develop such a system, we propose a global anti-intrusions system which detects and predicts intrusions based on hybrid propagation in Bayesian networks.

## IV. BAYESIAN NETWORKS

A Bayesian network is a graphical modeling tool used to model decision problems containing uncertainty. It is a directed acyclic graph where each node represents a discrete random variable of interest. Each node contains the states of the random variable that it represents and a conditional probability table (CPT) which give conditional probabilities of this variable such as realization of other connected variables, based upon Bayes rule:

$$\Pi(B|A)=\Pi(A|B)\Pi(B)/\Pi(A) \qquad (1)$$

The CPT of a node contains probabilities of the node being in a specific state given the states of its parents. The parent-child relationship between nodes in a Bayesian network indicates the direction of causality between the corresponding variables. That is, the variable represented by the child node is causally dependent on the ones represented by its parents [18].

Several researchers have been interested by using Bayesian network to develop intrusion detection systems. Axelsson in [5] wrote a well-known paper that uses the Bayesian rule of conditional probability to point out the implications of the *base-rate fallacy* for intrusion detection. It clearly demonstrates the difficulty and necessity of dealing with false alerts.

Kruegel in [1] presented a model that simulates an intelligent attacker using Bayesian techniques to create a plan of goal-directed actions. An event classification scheme is proposed based on Bayesian networks. Bayesian networks improve the aggregation of different model outputs and allow one to seamlessly incorporate additional information.

Johansen in [9] suggested that a Bayesian system which provides a solid mathematical foundation for simplifying a seemingly difficult and monstrous problem that today's Network IDS fail to solve. He added that Bayesian Network IDS should differentiate between attacks and the normal network activity by comparing metrics of each network traffic sample.

Bayesian networks learning has several advantages. First, it can incorporate prior knowledge and expertise by populating the CPTs. It is also convenient to introduce partial evidence and find the probability of unobserved variables. Second, it is capable of adapting to new evidence and knowledge by belief updates through network propagation.

*A. Bayesian Network Learning Algorithm*

Methods for learning Bayesian graphical models can be partitioned into at least two general classes of methods: constraint-based search and Bayesian methods. The constraint-based approaches [19] search the data for conditional independence relations from which it is in principle possible to deduce the Markov equivalence class of the underlying causal graph. Two notable constraint based algorithms are the PC algorithm which assumes that no hidden variables are present and the FCI algorithm which is capable of learning something about the causal relationships even assuming there are latent variables present in the data [19].

Bayesian methods [21] utilize a search-and-score procedure to search the space of DAGs, and use the posterior density as a scoring function. There are many variations on Bayesian methods, however, most research has focused on the application of greedy heuristics, combined with techniques to avoid local maxima in the posterior density (e.g., greedy search with random restarts or best first searches). Both constraint-based and Bayesian approaches have advantages and disadvantages. Constraint-based approaches are relatively quick and possess the ability to deal with latent variables. However, constraint-based approaches rely on an arbitrary significance level to decide independencies.

Bayesian methods can be applied even with very little data where conditional independence tests are likely to break down. Both approaches have the ability to incorporate background knowledge in the form of temporal ordering, or forbidden or forced arcs. Also, Bayesian approaches are capable of dealing with incomplete records in the database. The most serious drawback to the Bayesian approaches is the fact that they are relatively slow.

In this paper, we are dealing with incomplete records in the database so we opted for the Bayesian approach and particularly for the K2 algorithm. K2 learning algorithm showed high performance in many research works. The principle of K2 algorithm, proposed by Cooper and Herskovits, is to define a database of variables : X1,..., Xn, and to build an acyclic graph directed (DAG) based on the calculation of local score [22]. Variables constitute network nodes. Arcs represent "causal" relationships between variables.

K2 Algorithm used in learning step needs:

- A given order between variables
- and the number of parents of the node.

K2 algorithm proceeds by starting with a single node (the first variable in the defined order) and then incrementally adds connection with other nodes which can increase the whole probability of network structure, calculated using the S function. A requested new parent which does not increase node probability can not be added to the node parent set.

$$S(V_i, C(V_i)) = \prod_{j=1}^{q_i} \frac{(r_i - 1)!}{(N_{ij} + r_i - 1)!} \prod_{k=1}^{r_i} N_{ijk}! \quad (2)$$

Where, for each variable $V_i$, $r_i$ is the number of possible instantiations, $N_{ij}$ is the j-th instantiation of $C(V_i)$ in the database, $q_i$ is the number of possible instantiations for $C(V_i)$, $N_{ijk}$ is the number of cases in D for which $V_i$ takes the value $V_{ik}$ with $C(V_i)$ instantiated to $N_{ij}$, $N_{ij}$ is the sum of $N_{ijk}$ for all values of k.

## V. JUNCTION TREE INFERENCE ALGORITHM

The most common method to perform discrete exact inference is the Junction Tree algorithm developed by Jensen [23]. The idea of this procedure is to construct a data structure called a junction tree which can be used to calculate any query through message passing on the tree.

The first step of JT algorithm creates an undirected graph from an input DAG through a procedure called moralization. Moralization keeps the same edges, but drops the direction, and then connects the parents of every child. Junction tree construction follows four steps:

- *JT Inference Step1*: Choose a node ordering. Note that node ordering will make a difference in the topology of the generated tree. An optimal node ordering with respect to the junction tree is NP-hard to find.
- *JT Inference Step2*: Loop through the nodes in the ordering. For each node $X_i$, create a set $S_i$ of all its neighbours. Delete the node $X_i$ from the moralized graph.
- *JT Inference Step3*: Build a graph by letting each $S_i$ be a node. Connect the nodes with weighted undirected edges. The weight of an edge going from $S_i$ to $S_j$ is $|S_i \cap S_j|$.
- *JT Inference Step4*: Let the junction tree be the maximal-weight spanning tree of the cluster graph.

## VI. PROBLEM DESCRIPTION

The inference in bayesian networks is the post-calculation of uncertainty. Knowing the states of certain variables (called variables of observation), the inference process determines the states of some other variables (called variables targets) conditionally to observations. The choice to represent the knowledge by probabilities only, and therefore to suppose that the uncertainty of the information we dispose has stochastic origin, has repercussions on the results of uncertainty propagation through the bayesian model.

The two sources of uncertainties (stochastic - epistemic) must be treated in different manners. In practice, while the uncertain information is treated with rigorous manner by the classic probability distributions, the imprecise information is much more better treated by possibility distribution. The two sources of uncertainty don't exclude themselves and are often related (for example: imprecise measurement of an uncertain quantity ).

The merely probabilistic propagation approach can generate some too optimistic results. This illusion is reinforced by the fact that information is sometimes imprecise or incomplete and the classic probabilistic context doesn't represent this type of information faithfully.

In the section below, we will present a new propagation approach in bayesian networks called hybrid propagation combining probabilities and possibilities. The advantage of this approach over the classic probabilistic propagation is that it takes into account both the uncertain and the imprecise character of information.

## VII. HYBRID PROPAGATION IN BAYESIAN NETWORKS

The mechanism of propagation is based on Bayesian model. Therefore, the junction tree algorithm is used for the inference in the Bayesian network. The hybrid calculation combining probabilities and possibilities, permits to propagate both the variability (uncertain information) and the imprecision (imprecise information).

### A. Variable Transformation from Probability to Possibility (TV)

Let's consider the probability distribution p=(p₁,...,pᵢ,...,pₙ) ordered as follows: p₁>p₂>…>pₙ. The possibility distribution π=(π₁,…,πᵢ,…,πₙ) according to the transformation (p→π) proposed in [24] is π₁>π₂>…>πₙ. Every possibility is defined by:

$$\pi_i = \left(\frac{p_i}{p_1}\right)^{k_i(1-p_i)} \quad \forall \ i = 1, 2, .., n \qquad (3)$$

Where $k_1$=1, $k_i = \frac{\log (p_i+p_{i+1}+\cdots p_n)}{(1-p_i).\log (\frac{p_i}{p_1})}$, $\forall$ i =2, 3, …, n

### B. Probability Measure and Possibility Distribution

Let's consider a probabilistic space (Ω,A,P). For all measurable whole A⊆Ω, we can define its high probability and its low probability. In other terms the value of the probability P(A) is imprecise: $\forall$ A⊆Ω, N(A) ≤ P(A) ≤ Π(A) where N(A) = 1-Π($\bar{A}$).

Each couple of necessity/possibility measures (N,Π) can be considered as the lower and higher probability measures induced by a probability measure. The gap between these two measures reflects the imprecise character of the information. It is about defining a possibility distribution on a probability measure. This possibility distribution reflects the imprecise character of the true probability of the event.

A probability measure is more reliable and informative when the gap between its two upper and lower terminals is reduced, ie imprecision on the value of the variable is reduced, as opposed to a measure of probability in a confidence interval relatively large, this probability is risky and not very informative.

### C. Hybrid Propagation Process

The hybrid propagation proceeds in three steps:

1) Substitute probability distributions of each variable in the graph by probability distributions framed by measures of possibility and necessity, using the variable transformation from probability to possibility TV, applied to probability distributions of each variable in the graph. The gap between the necessity and possibility measures reflects the imprecise character of the true probability associated to the variable.

2) Transformation of the initial graph to a junction tree.

3) Uncertain and imprecise uncertainty propagation which consists in both :

   a) The classic probabilistic propagation of stochastic uncertainties in junction tree through message passing on the tree, and

   b) The possibilistic propagation of epistemic uncertainties in the junction tree. Possibilistic propagation in junction tree is a direct adaptation of the classic probabilistic propagation.

Therefore, the proposed propagation method:

1) Preserves the power of modeling of Bayesian networks (permits the modeling of relations between variables),

2) This method is adapted to both stochastic and epistemic uncertainties,

3) The result is a probability described by an interval delimited by possibility and necessity measures.

## VIII. HYBRID INTRUSION DETECTION AND PREDICTION SYSTEM

Our anti-intrusions system operates to two different levels, it integrates host intrusion detection and network intrusion prediction.

The intrusion detection consists in analyzing audit data of the host in search of attacks whose signatures are stocked in a signatures dataset. The intrusion prediction, consists in analyzing the stream of alerts resulting from one or several detected attacks, in order to predict the possible attacks that will follow in the whole network.

Our anti-intrusions approach is based on hybrid propagation in Bayesian networks in order to benefit the power of modeling of Bayesian networks and the power of possibilistic reasoning to manage imprecision.

### A. Hybrid intrusion detection

The main objective of intrusion detection is to detect each security policy violation on a system of information. Signature Recognition approach, adopted in our contribution, analyzes audit data in search of attacks whose signatures are stocked in a signatures dataset. Audit data are data of the computer system that bring back information on operations led on this later. A signatures dataset contains a set of lines, every line codes a stream of data (between two definite instants) between a source (identified by its IP address) and a destination (identified also by its IP address), under a given protocol (TCP, UDP...). Every line is a connection characterized by a certain number of attributes as its length, the type of the protocol, etc. According to values of these attributes, every connection in the signatures dataset is considered as being a normal connection or an attack.

In our approach, the process of intrusion detection is considered as a problem of classification. Given a set of identified connections and a set of connections types, our goal is to classify connections among the most plausible corresponding connections types.

Our approach for intrusion detection consists in four main steps [30]:

*1)* *Important attributes selection :* In a signatures dataset, every connection is characterized by a certain number of attributes as its length, the type of the protocol, etc. These attributes have been fixed by Lee and al. [31]. The objective of this step is to extract the most important attributes among attributes of the signatures dataset. To do so, we proceeded by a Multiple Correspondences Factorial Analysis (MCFA) of attributes of the dataset, then we calculated the Gini index for every attribute of the dataset in order to visualize the different attributes distribution and to select the most important attributes [32]. It results a set of the most important attributes characterizing connections of the signatures dataset. Some of these attributes can be continuous and can require to be discretized to improve classification results,

*2)* *Continuous attributes discretization :* The selected attributes, can be discrete (admitting a finished number of values) or continuous. Several previous works showed that the discretization improved bayesian networks performances [4]. To discretize continuous attributes, we opted for the discretization by the fit together averages. it consists in cutting up the variable while using some successive averages as limits of classes. This method has the advantage to be bound strongly to the variable distribution, but if the variable is cut up a big number of time, this method risks to either produce some empty or very heterogeneous classes, in the case of very dissymmetric distributions. Thus, we use only one iteration, i.e. a binary discretization based on the average, but this supposes that the behavior of the observation variables is not too atypical,

*3)* *Bayesian network learning :* The set of important attributes being discretized as well as the class of connection types constitute the set of entry variables to the Bayesian network learning step. The first step is to browse the set of entry variables to extract their different values and to calculate their probabilities. Then, we use the K2 probabilistic learning algorithm to build the Bayesian network for intrusion detection. The result is a directed acyclic graph whose nodes are the entry variables and edges denote the conditional dependences between these variables. To each variable of the graph is associated a conditional probability table that quantifies the effect of its parents,

*4)* *Hybrid propagation in Bayesian network :* consists in the three steps mentioned previously.

At the end of this step, every connection (normal or intrusion) in a host is classified in the most probable connection type. In case of detected intrusions in a host, one or several alerts are sent in direction of the intrusion prediction module, this later is charged to predict the possible intrusions that can propagate in the whole network.

### B. Hybrid intrusion prediction

The intrusion prediction aims to predict attack plans, given one or several intrusions detected at the level of one or several hosts of a computer network. An intrusion detected at the level of a host results to one or several alerts generated by a HIDS. The intrusion prediction tent to classify alerts among the most plausible hyper-alerts, each hyper-alert corresponds to a step in the attack plan, then, based on hyper-alerts correlation, deducts the possible attacks that will follow in the whole computer network [11].

*1)* *Alerts Classification:* The main objective of the alerts classification is to analyze the stream of alerts generated by intrusion detection systems in order to contribute in the attack plans prediction. In our approach, given a set of alerts, alerts classification's goal is to classify alerts among the most plausible corresponding hyper-alerts.

Our approach for alerts classification consists in four main steps :

*a)* *Important attributes selection:* In addition to time information, each alert has a number of other attributes, such as source IP, destination IP, port(s), user name, process name, attack class, and sensor ID, which are defined in a standard document, "Intrusion Detection Message Exchange Format (IDMEF)", drafted by the IETF Intrusion Detection Working Group [20]. For the most important attributes selection, we proceed by a Multiple Correspondences Factorial Analysis (MCFA) of the different attributes characterizing alerts. The attributes selection doesn't include time stamps, we will use time stamps in the attack plans prediction process in order to detect alerts series. It results of this step, a set of the most important attributes characterizing alerts of the alerts dataset,

*b)* *Alerts aggregation:* An alerts dataset generally contains a big number of alerts, most are raw alerts and can make reference to one same event. Alerts aggregation consists in exploiting alerts attributes similarities in order to reduce the redundancy of alerts. Since alerts that are output by the same IDS and have the same attributes except time stamps correspond to the same step in the attack plan [26], we aggregate alerts sharing the same sensor, the same attributes except time stamps in order to get clusters of alerts where each cluster corresponds to only one step of the attack plan, called hyper-alert. Then, based on results of this first step, we merge clusters of alerts (or hyper-alert) corresponding to the same step of the attack plan. At the end of this step of alerts aggregation, we get, a cluster of alerts (or hyper-alert) for each step of the attack plan (i.e. hyper-alert = step of the attack plan). We regroup in one class all the observed hyper-alerts,

*c) Bayesian network learning:* The set of selected attributes of alerts as well as the class regrouping all the observed hyper-alerts forms the set of entry variables to the Bayesian network learning step. The first step is to browse the set of entry variables in order to extract their different values and calculate their probabilities. Then, we use the K2 probabilistic learning algorithm to build the Bayesian network for alerts classification,

*d) Hybrid propagation in Bayesian network :* consists in the three steps mentioned previously.

At the end of this step, every generated alert is classified in the most probable corresponding hyper-alert.

2) *Attack plans prediction:* Attack plans prediction consists in detecting complex attack scenarios, that is implying a series of actions by the attacker. The idea is to correlate hyper-alerts resulting from the previous step in order to predict, given one or several attacks detected, the possible attacks that will follow.

Our approach for attack plans prediction consists in three main steps :

a) *Transaction data formulation [26]:* we formulate transaction data for each hyper alert in the dataset. Specifically, we set up a series of time slots with equal time interval, denoted as $\Delta t$, along the time axis. Given a time range T, we have $m = \frac{T}{\Delta t}$ time slots. Recall that each hyper alert A includes a set of alert instances with the same attributes except time stamps, i.e., $A = [a_1, a_2, ..., a_n]$, where $a_i$ represents an alert instance in the cluster. We denote $N_A = \{n_1, n_2, ..., n_m\}$ as the variable to represent the occurrence of hyper alert A during the time range T, where $n_i$ is corresponding to the occurrence (i.e., $n_i = 1$) or un-occurrence (i.e., $n_i = 0$) of the alert A in a specific time slot $\Delta t_i$, Using the above process, we can create a set of transaction data. Table 1 shows an example of the transaction data corresponding to hyper alerts *A*, *B* and *C*.

TABLE I.        AN EXAMPLE OF TRANSACTION DATA SET

| Time Slot | A | B | C |
|---|---|---|---|
| $\Delta t_1$ | 1 | 0 | 0 |
| $\Delta t_2$ | 1 | 0 | 1 |
| $\Delta t_3$ | 1 | 1 | 0 |
| ... | ... | ... | ... |
| $\Delta t_m$ | 1 | 0 | 0 |

b) *Bayesian network learning:* The set of the observed hyper-alerts forms the set of entry variables to the Bayesian network learning step. The first step is to browse the set of entry variables to extract their different values and to calculate their probabilities. Then, we use the K2 probabilistic learning algorithm to build the Bayesian network for attack plans prediction. The result is a directed acyclic graph whose nodes are the hyper-alerts and edges denote the conditional dependences between these hyper-alerts.

*c) Hybrid propagation in Bayesian network :* consists in the three steps mentioned previously.

At the end of this step, given one or several attacks detected, we can predict the possible attacks that will follow.

## IX.    HIDPAS SYSTEM AGENT ARCHITECTURE

HIDPAS system architecture is composed by two interconnected layers of intelligent agents. The first layer is concerned by host intrusion detection. On each host of a distributed computers system an intelligent agent is charged by detecting intrusion eventuality.

Each agent of the intrusion detection layer uses a signature intrusion database (SDB) to build its own bayesian network. For every new suspect connection, the intrusion detection agent (IDA) of the concerned host uses hybrid propagation in its bayesian network to infer the conditional evidences of intrusion given the new settings of the suspect connection. Therefore, based on the probability degree and the gap between the necessity and the possibility degrees associated with each connection type, we can perform quantitative analysis on the connection types.

In the final selection of possible connection type, we can select the type who has the maximum informative probability value. An informative probability is a probability delimited by two confidence measures where the gap between them is under a threshold.

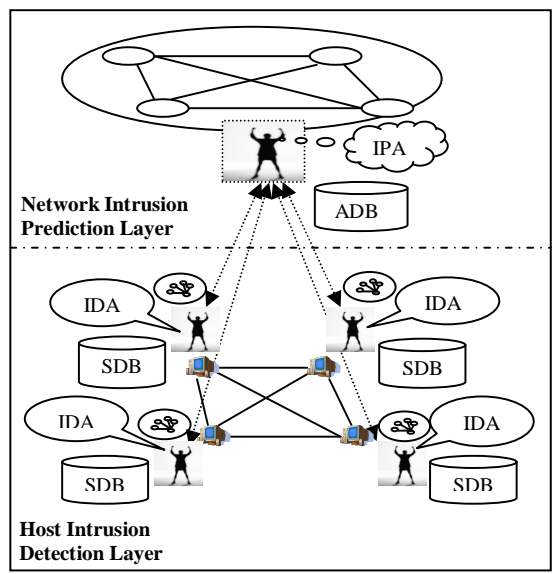

Figure 1.   HIDPAS system architecture

In case of intrusion, IDA agent informs the intrusion prediction agent (IPA) which is placed in the prediction layer,

about the eventuality of intrusion on the concerned host and its type.

The second layer is based upon one intelligent agent which is charged by network intrusion prediction.

When the Intrusion Prediction Agent (IPA) is informed about a new intrusion which will be happened on a host of the distributed computers system and its type, it tries to compute conditional probabilities that other attacks may be ultimately happen. To accomplish this task, IPA uses another database type (ADB) which contains historical data about alerts generated by sensors from different computer systems.

Given a stream of alerts, agent IPA first output results as evidences to the inference process of the first graph for alerts classification, second, it output results of alerts classification to the inference process of the second graph for attack plans prediction. Each path in the second graph is potentially a subsequence of an attack scenario. Therefore, based on the probability degree and the gap between the necessity and the possibility degrees associated with each edge, IPA can perform quantitative analysis on the attack strategies.

The advantage of our approach is that we do not require a complete ordered attack sequence for inference. Due to bayesian networks and the hybrid propagation, we have the capability of handling partial order and unobserved activity evidence sets. In practice, we cannot always observe all of the attacker's activities, and can often only detect partial order of attack steps due to the limitation or deployment of security sensors. For example, IDA can miss detecting intrusions and thus result in an incomplete alert stream.

In the final selection of possible future goal or attack steps, IPA can either select the node(s) who has the maximum informative probability value(s) or the one(s) whose informative probability value(s) is (are) above a threshold.

After computing conditional probabilities of possible attacks, IPA informs the system administrator about possible attacks.

## X. HIDPAS SYSTEM IMPLEMENTATION

HIDPAS was implemented using JADE multiagent plateform. The dataset used for intrusion detection implementation and experimentation is DARPA KDD'99 which contains signatures of normal connections and signatures of 38 known attacks gathered in four main classes: DOS, R2L, U2R and Probe.

### A. DARPA'99 DATA SET

MIT Lincoln Lab's DARPA intrusion detection evaluation datasets have been employed to design and test intrusion detection systems. The KDD 99 intrusion detection datasets are based on the 1998 DARPA initiative, which provides designers of intrusion detection systems (IDS) with a benchmark on which to evaluate different methodologies [25].

To do so, a simulation is made of a factitious military network consisting of three 'target' machines running various operating systems and services. Additional three machines are then used to spoof different IP addresses to generate traffic.

Finally, there is a sniffer that records all network traffic using the TCP dump format. The total simulated period is seven weeks [27]. Packet information in the TCP dump file is summarized into connections. Specifically, "a connection is a sequence of TCP packets starting and ending at some well defined times, between which data flows from a source IP address to a target IP address under some well defined protocol" [27].

DARPA KDD'99 dataset represents data as rows of TCP/IP dump where each row consists of computer connection which is characterized by 41 features. Features are grouped into four categories:

1) Basic Features: Basic features can be derived from packet headers without inspecting the payload.

2) Content Features: Domain knowledge is used to assess the payload of the original TCP packets. This includes features such as the number of failed login attempts;

3) Time-based Traffic Features: These features are designed to capture properties that mature over a 2 second temporal window. One example of such a feature would be the number of connections to the same host over the 2 second interval;

4) Host-based Traffic Features: Utilize a historical window estimated over the number of connections – in this case 100 – instead of time. Host based features are therefore designed to assess attacks, which span intervals longer than 2 seconds.

In this study, we used KDD'99 dataset which is counting almost 494019 of training connections. Based upon a Multiple Correspondences Factorial Analysis (MCFA) of attributes of the KDD'99 dataset, we used data about only important features. Table 2 shows the important features and the corresponding Gini index for each feature:

TABLE II. CONNECTIONS IMPORTANT FEATURES

| N° | Feature | Gini |
|-----|-------------------------------|--------|
| A23 | Count | 0.7518 |
| A5 | src_bytes | 0.7157 |
| A24 | src_count | 0.6978 |
| A3 | service | 0.6074 |
| A36 | dst_host_same_src_port_rate | 0.5696 |
| A2 | protocol_type | 0.5207 |
| A33 | dst_host_srv_count | 0.5151 |
| A35 | dst_host_diff_srv_rate | 0.4913 |
| A34 | dst_host_same_srv_rate | 0.4831 |

To these features, we added the "attack_type". Indeed each training connection is labelled as either normal, or as an attack with specific type. DARPA'99 base counts 38 attacks which can be gathered in four main categories:

1) Denial of Service (dos): Attacker tries to prevent legitimate users from using a service.

2) **Remote to Local (r2l):** Attacker does not have an account on the victim machine, hence tries to gain access.

3) **User to Root (u2r):** Attacker has local access to the victim machine and tries to gain super user privileges.

4) **Probe:** Attacker tries to gain information about the target host.

Among the selected features, only service and protocol-type are discrete, the other features need to be discretized. Table 3 shows the result of discretization of these features.

TABLE III.    CONTINUOUS FEATURES DISCRETIZATION

| N° | Feature | Values |
|----|---------|--------|
| A23 | count | cnt_v1<br>m < 332.67007446<br>cnt_v2<br>m ≥ 332.67007446 |
| A5 | src_bytes | sb_v1<br>m < 30211.16406250<br>sb_v2<br>m ≥ 30211.16406250 |
| A24 | src_count | srv_cnt_v1<br>m < 293.24423218<br>srv_cnt_v2<br>m ≥ 293.24423218 |
| A36 | dst_host | dh_ssp_rate_v1<br>m < 0.60189182<br>dh_ssp_rate_v2<br>m ≥ 0.60189182 |
| A33 | dst_host_srv_count | dh_srv_cnt_v1<br>m < 189.18026733<br>dh_srv_cnt_v2<br>m ≥ 189.18026733 |
| A35 | dst_host_diff_srv_rate | dh_dsrv_rate_v1<br>m < 0.03089163<br>dh_dsrv_rate_v2<br>m ≥ 0.03089163 |
| A34 | dst_host_same_srv_rate | dh_ssrv_rate_v1<br>m < 0.75390255<br>dh_ssrv_rate_v2<br>m ≥ 0.75390255 |

The dataset used for intrusion prediction implementation and experimentation is LLDOS 1.0 provided by DARPA 2000, which is the first attack scenario example dataset to be created for DARPA. It includes a distributed denial of service attack run by a novice attacker.

### B.  LLDOS 1.0 – SCENARIO ONE

DARPA2000 is a well-known IDS evaluation dataset created by the MIT Lincoln Laboratory. It consists of two multistage attack scenarios, namely LLDDOS1.0 and LLDOS2.02. The LLODS1.0 scenario can be divided into five phases as follows [29]:

1) **Phase 1:** The attacker scans the network to determine which hosts are up.

2) **Phase 2:** The attacker then uses the *ping* option of the *sadmind* exploit program to determine which hosts selected in Phase 1 are running the Sadmind service.

3) **Phase 3:** The attacker attempts the sadmind Remote-to-Root exploit several times in order to compromise the vulnerable machine.

4) **Phase 4:** The attacker uses *telnet* and *rpc* to install a DDoS program on the compromised machines.

5) **Phase 5:** The attacker telnets to the DDoS master machine and launches the *mstream DDOS* against the final victim of the attack.

We used an alert log file [28] generated by RealSecure IDS. As a result of replaying the "Inside-tcpdump" file from DARPA 2000, Realsecure produces 922 alerts. After applying the proposed alerts important attributes selection, we used data about only important features as shown in Table4.

TABLE IV.    ALERTS IMPORTANT FEATURES

| Feature | Gini |
|---------|------|
| SrcIPAddress | 0,6423 |
| SrcPort | 0,5982 |
| DestIPAddress | 0,5426 |
| DestPort | 0,5036 |
| AttackType | 0,4925 |

After applying the proposed alerts aggregation, we obtained 17 different types of alerts as shown in Table5.

TABLE V.    HYPER-ALERTS REPORTED BY REALSECURE IN LLDOS 1.0

| ID | Hyper-alert | Size |
|----|-------------|------|
| 1 | Sadmind_Ping | 3 |
| 2 | TelnetTerminaltype | 128 |
| 3 | Email_Almail_Overflow | 38 |
| 4 | Email_Ehlo | 522 |
| 5 | FTP_User | 49 |
| 6 | FTP_Pass | 49 |
| 7 | FTP_Syst | 44 |
| 8 | http_Java | 8 |
| 9 | http_Shells | 15 |
| 10 | Admind | 17 |
| 11 | Sadmind_Amslverify_Overflow | 14 |
| 12 | Rsh | 17 |
| 13 | Mstream_Zombie | 6 |
| 14 | http_ Cisco_ Catalyst_ Exec | 2 |
| 15 | SSH_Detected | 4 |
| 16 | Email_Debug | 2 |
| 17 | Stream_DoS | 1 |

### C.  SYSTEM IMPLEMENTATION

HIDPAS system contains three interfaces:

1) **Intrusion Detection Interface :** Figure 2 shows the bayesian network built by AGENT ID1. For every new connection, AGENT ID1 uses its bayesian network to decide about the intrusion and its type.

2) **Alerts Classification Interface :** Figure 3 shows the bayesian network built by the IPA for alerts classification. The IPA receives alerts messages sent by intrusion detection agents about the detected intrusions. The IPA uses its bayesian network to determine hyper-alerts corresponding to these alerts.

3) **Attack Plans Prediction Interface :** Figure 4 shows the bayesian network built by the IPA for attack plans prediction.

The IPA uses its bayesian network to determine the eventual attacks that will follow the detected intrusions.

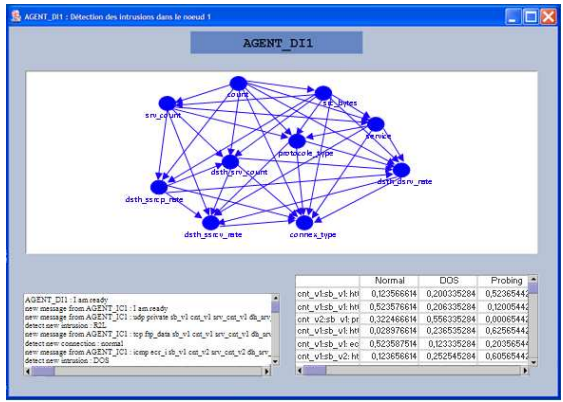

Figure 2.  Intrusion Detection Interface

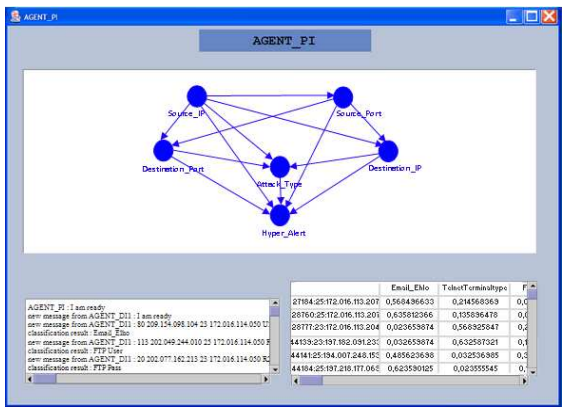

Figure 3.  Alerts Classification Interface

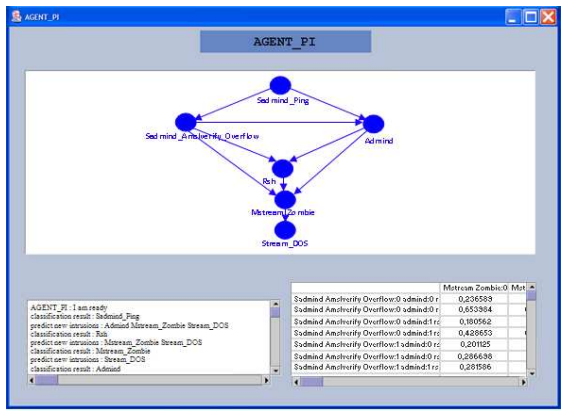

Figure 4.  Intrusion Prediction Interface

## XI.  EXPERIMENTATION

The main criteria that we have considered in the experimentation of our system are the detection rate, false alerts rate, alerts correlation rate, false positive correlation rate and false negative correlation rate.

- Detection Rate: is defined as the number of examples correctly classified by our system divided by the total number of test examples.

TABLE VI.  DETECTION RATE COMPARISON

| Detection | Classic propagation | Hybrid propagation |
|---|---|---|
| **Normal** (60593) | 99.52% | 100% |
| **DOS** (229853) | 97.87% | 99.93% |
| **Probing** (4166) | 89.39% | 98.57% |
| **R2L** (16189) | 19.03% | 79.63% |
| **U2R** (228) | 31.06% | 93.54% |

Table 6 shows high performance of our system based on hybrid propagation in intrusion detection.

- False Alerts : Bayesian networks can generate two types of false alerts: False negative and false positive alarms. *False negative* describe an event that the IDS fails to identify as an intrusion when one has in fact occurred. *False positive* describe an event, incorrectly identified by the IDS as being an intrusion when none has occurred.

TABLE VII.  FALSE ALERTS RATE COMPARISON

| False alerts | Classic propagation | Hybrid propagation |
|---|---|---|
| **Normal** (60593) | 0.48% | 0% |
| **DOS** (229853) | 1.21% | 0.02% |
| **Probing** (4166) | 5.35% | 0.46% |
| **R2L** (16189) | 6.96% | 2.96% |
| **U2R** (228) | 6.66% | 1.36% |

Table 7 shows the gap between false alerts results given by the two approaches. Hybrid propagation approach gives the smallest false alerts rates.

- Correlation rate: can be defined as the rate of attacks correctly correlated by our system.

- False positive correlation rate : is the rate of attacks correlated by the system when no relationship exists between them.

- False negative correlation rate : is the rate of attacks having in fact relationship but the system fails to identify them as correlated attacks.

Table 8 shows experimentation results about correlation measured by our system:

TABLE VIII.  CORRELATION RATE COMPARISON

| | Classic propagation | hybrid propagation |
|---|---|---|
| **Correlation rate** | 95.5% | 100% |
| **False positive correlation rate** | 6.3% | 1.3% |
| **False negative correlation rate** | 4.5% | 0% |

Table 8 shows high performance of our system based on hybrid propagation in attack correlation and prediction. The use of hybrid propagation in bayesian networks was especially useful, because we have deal with a lot of missing information.

## XII. CONCLUSION

In this paper, we outlined a new approach based on hybrid propagation combining probability and possibility, through a bayesian network. Bayesian networks provide automatic learning from audit data. Hybrid propagation through bayesian network provide propagation of both stochastic and epistemic uncertainties, coming respectively from the uncertain and imprecise character of information.

The application of our system in intrusion detection context helps detect both normal and abnormal connections with very considerable rates.

Besides, we presented an approach to identify attack plans and predict upcoming attacks. We developed a bayesian network based system to correlate attack scenarios based on their relationships. We conducted inference to evaluate the likelihood of attack goal(s) and predict potential upcoming attacks based on the hybrid propagation of uncertainties.

Our system demonstrates high performance when detecting intrusions, correlating and predicting attacks. This is due to the use of bayesian networks and the hybrid propagation within bayesian networks which is especially useful when dealing with missing information.

There are still some challenges in attack plan recognition. First, we will apply our algorithms to alert streams collected from live networks to improve our work. Second, our system can be improved by integrating an expert system which is able to provide recommendations based on attack scenarios prediction.

### REFERENCES

[1] Kruegel Christopher, Darren Mutz William, Robertson Fredrik Valeur. Bayesian Event Classification for Intrusion Detection Reliable Software Group University of California, Santa Barbara, , 2003.

[2] F. Cuppens and R. Ortalo. LAMBDA: A language to model a database for detection of attacks. In Third International Workshop on the Recent Advances in Intrusion Detection (RAID'2000), Toulouse, France, 2000.

[3] C. Baudrit and D. Dubois. Représentation et propagation de connaissances imprécises et incertaines: Application à l'évaluation des risques liées aux sites et aux sols pollués. Université Toulouse III – Paul Sabatier, Toulouse, France, Mars 2006.

[4] Dougherty J., Kohavi R., Sahami M., «Forrests of fuzzy decision trees», Proceedings of ICML'95 : supervised and unsupervised discretization of continuous features, p. 194-202, 1995.

[5] S. Axelsson. The Base-Rate Fallacy and its Implications for the Difficulty of Intrusion Detection. In 6th ACM Conference on Computer and Communications Security, 1999.

[6] Christopher Kruegel, Darren Mutz William, Robertson Fredrik Valeur, Bayesian Event Classification for Intrusion Detection Reliable Software Group University of California, Santa Barbara, 2003.

[7] R. Goldman. A Stochastic Model for Intrusions. In Symposium on Recent Advances in Intrusion Detection (RAID), 2002.

[8] A. Valdes and K. Skinner. Adaptive, Model-based Monitoring for Cyber Attack Detection. In Proceedings of RAID 2000, Toulouse, France, October 2000.

[9] Krister Johansen and Stephen Lee, Network Security: Bayesian Network Intrusion Detection (BNIDS) May 3, 2003.

[10] Surya Kumari Govindu, Intrusion Forecasting System, http://www.securitydocs.com/library/3110 15/03/2005.

[11] Jemili F., Zaghdoud M., Ben Ahmed M., « Attack Prediction based on Hybrid Propagation in Bayesian Networks », In Proc. of the Internet Technology And Secured Transactions Conference, ICITST-2009.

[12] B. Landreth, Out of the Inner Circle, A Hacker's Guide to Computer Security, Microsoft Press, Bellevue, WA, 1985.

[13] Paul Innella and Oba McMillan, An introduction to Intrusion detection, Tetrad Digital Integrity, LLC, December 6, 2001, by URL: http://www.securityfocus.com, 2001.

[14] Brian C. Rudzonis, Intrusion Prevention: Does it Measure up to the Hype? SANS GSEC Practical v1.4b. April 2003.

[15] M. Roesch. Snort - Lightweight Intrusion Detection for Networks. In USENIX Lisa 99, 1999.

[16] K. Ilgun. USTAT: A Real-time Intrusion Detection System for UNIX. In Proceedings of the IEEE Symposium on Research on Security and Privacy, Oakland, CA, May 1993.

[17] Biswanath Mukherjee, Todd L. Heberlein and Karl N. Levitt, Network intrusion detection. IEEE Network, 8(3):26{41, May/June 1994.

[18] Peter Spirtes, Thomas Richard-son, and Christopher Meek. Learning Bayesian networks with discrete variables from data. In Proceedings of the First International Conference on Knowledge Discovery and Data Mining, pages 294-299, 1995.

[19] Peter Spirtes, Clark Glymour, and Richard Scheines. Causation, Prediction, and Search. Springer Verlag, New York, 1993.

[20] GROUP, I. I. D. W., « Intrusion Detection Message Exchange Format » http://www.ietf.org/internet-drafts/draft-ietf-idwg-idmef-xml-09.txt, 2002.

[21] Gregory F. Cooper and Edward Herskovits. A Bayesian method for the induction of probabilistic networks from data. Machine Learning, 1992.

[22] Sanguesa R., Cortes U. Learning causal networks from data: a survey and a new algorithm for recovering possibilistic causal networks. AI Communications 10, 31–61, 1997.

[23] Frank Jensen, Finn V. Jensen and Soren L. Dittmer. From influence diagrams to junction trees. Proceedings of UAI, 1994.

[24] M. Sayed Mouchaweh, P. Bilaudel and B. Riera. "Variable ProbabilityPossibility Transformation", 25th European Annual Conference on Human Decision-Making and Manual Control (EAM'06), September 27-29,Valenciennes, France, 2006.

[25] DARPA. Knowledge Discovery in Databases, 1999. DARPA archive. Task Description http://www.kdd.ics.uci.edu/databases/kddcup99/task.htm

[26] Qin Xinzhou, «A Probabilistic-Based Framework for INFOSEC Alert Correlation », PhD thesis, College of Computing Georgia Institute of Technology, August 2005.

[27] Kayacik, G. H., Zincir-Heywood, A. N. Analysis of Three Intrusion Detection System Benchmark Datasets Using Machine Learning Algorithms, Proceedings of the IEEE ISI 2005 Atlanta, USA, May, 2005.

[28] North Carolina State University Cyber Defense Laboratory, *Tiaa: A toolkit for intrusion alert analysis*, http://discovery.csc.ncsu.edu/software/correlator/ver0.4/index.html

[29] MIT Lincoln Laboratory, *2000 darpa intrusion detection scenario speciˆc data sets*, 2000.

[30] Jemili F., Zaghdoud M., Ben Ahmed M., « Intrusion Detection based on Hybrid Propagation in Bayesian Networks », In Proc. of the IEEE International Conference on Intelligence and security informatics, ISI 2009.

[31] Lee W., Stolfo S. J., Mok K. W., « A data mining framework for building intrusion detection models », Proceedings of the 1999 IEEE symposium on security and privacy, 1999.

[32] Arfaoui N., Jemili F., Zaghdoud M., Ben Ahmed M., « Comparative Study Between Bayesian Network And Possibilistic Network In Intrusion Detection », In Proc. of the International Conference on Security and Cryptography, Secrypt 2006.